\documentclass[letterpaper,12pt]{article}
\pdfoutput=1
\usepackage{jheppub}
\usepackage{graphics,epsfig}
\usepackage{graphicx}
\usepackage{amsmath}
\usepackage{amssymb}
\usepackage{bbm}
\usepackage{subfigure}
\usepackage{epsfig}
\usepackage{yfonts}
\usepackage{braket} 
\usepackage{lipsum} 

\newcommand{\be}{\begin{equation}}
\newcommand{\ee}{\end{equation}}
\newcommand{\bea}{\begin{eqnarray}}
\newcommand{\eea}{\end{eqnarray}}
\newcommand{\ba}{\begin{eqnarray}}
\newcommand{\ea}{\end{eqnarray}}

\newcommand{\beq}{\begin{equation}}
\newcommand{\eeq}{\end{equation}}
\newcommand{\beqa}{\begin{eqnarray}}
\newcommand{\eeqa}{\end{eqnarray}}
\newcommand{\beqar}{\begin{eqnarray*}}
\newcommand{\eeqar}{\end{eqnarray*}}

\newcommand{\reef}[1]{(\ref{#1})}

\newcommand{\tr}{{\rm tr}}

\def\({\left(} \def\){\right)}
\def\[{\left[} \def\]{\right]}

\title{Modular energy inequalities from relative entropy}

\author[a]{David Blanco}
\author[b]{, Horacio Casini}
\author[a]{, Mauricio Leston}
\author[a,c]{, Felipe Rosso}

\affiliation[a]{Instituto de Astronom\'ia y F\'isica del Espacio, Universidad Nacional de Buenos Aires,
(1428) Ciudad Aut\'onoma de Buenos Aires, Argentina.}
\affiliation[b]{Centro At\'omico Bariloche, (8400) S.C. de Bariloche, R\'{\i}o Negro, Argentina.}
\affiliation[c]{Department of Physics and Astronomy, University of Southern California, Los Angeles, CA 90089-0484, U.S.A.}
\emailAdd{dblanco@iafe.uba.ar}
\emailAdd{casini@cab.cnea.gov.ar}
\emailAdd{mauricio@iafe.uba.ar}
\emailAdd{felipero@usc.edu}

\abstract{We obtain new constraints for the modular energy of general states by using the monotonicity property of relative entropy. In some cases, modular energy can be related to the energy density of states and these constraints lead to interesting relations between energy and entropy. In particular, we derive new quantum energy inequalities that improve some previous bounds for the energy density of states in a conformal field theory. Additionally, the inequalities derived in this manner also lead us to conclude that the entropy of the state further restricts the possible amount of negative energy allowed by the theory.}

\begin{document}
\maketitle


\section{Introduction}

Relative entropy $S\left(\rho_1|\rho_0\right)$ between two states $\rho_0$ and $\rho_1$ in the same Hilbert space,
\begin{equation}
S\left(\rho_1|\rho_0\right)=\tr \left(\rho_1\log\rho_1\right)-\tr \left(\rho_1\log\rho_0\right)\,,
\end{equation}
is a fundamental concept in quantum information theory. This quantity gives us an operational definition of distinguishability between two states in the following sense: given a state $\rho_1$ the probability of confounding it with another state $\rho_0$ after $n$ trials of some measurement decays exponentially as $e^{-n S\left(\rho_0|\rho_1\right)}$ for large $n$ \cite{vedral}.

In contrast with the entanglement entropy of a state $\rho$ reduced to a spatial region $V$,
\begin{equation}
S(\rho_V)=-\tr\left(\rho_V \log\left(\rho_V\right)\right)\,,
\end{equation}
relative entropy is free from divergences in quantum field theory. This is due to the subtraction of the contributions coming from the entanglement between the high energy modes inside and outside $V$ localised around the border $\partial V$.

A key property of relative entropy is its \textit{positivity}, i.e.
\begin{equation}
S\left(\rho_1|\rho_0\right) \geq 0\,,\label{pos}
\end{equation}
for all states $\rho_0$ and $\rho_1$, where the equality only ocurrs when $\rho_1=\rho_0$. This property is equivalent to the fact that, at a fixed temperature $T$, the free energy $F\left(\rho\right)=\tr \left(\rho H\right)-T S\left(\rho\right)$ \footnote{$H$ stands for the dynamical hamiltonian of the system.} of a state $\rho$ is minimal for the Gibbs thermal state $\rho_T=\frac{e^{-H/T}}{\tr\left(e^{-H/T}\right)}$ of the system at that temperature, i.e., $F\left(\rho\right)\geq F\left(\rho_T\right)$ for arbitrary $\rho$.

Equation \reef{pos} has proven to be of wide use in a variety of topics. In particular, it is a key ingredient to establish a precise formulation of the Bekenstein bound \cite{beke} and the quantum Bousso bound \cite{bousso}. The positivity of relative entropy also plays an essential role in the proof of the first law of entanglement \cite{first-law}.

Relative entropy decreases under an arbitrary Completely Positive Trace-Preserving (CPTP) map $\Phi$, i.e.
\begin{equation}
S\left(\rho_1|\rho_0\right)\geq S\left(\Phi\left(\rho_1\right)|\Phi\left(\rho_0\right)\right)\,.\label{monotonicity}
\end{equation}
The second law of thermodynamics is intimately related to this property. In the canonical system, for instance, if we consider a CPTP map $\Phi$ that preserves the mean value of the energy and that keeps invariant the Gibbs thermal state $\rho_T$, it is straightforward to show that the entanglement entropy of the system does not decrease with the evolution under $\Phi$, i.e., that $S\left(\rho\right)\leq S\left(\Phi\left(\rho\right)\right)$.

If we consider states reduced to some spatial regions $A$ and $B$, with $B\subseteq A$, from equation \reef{monotonicity} follows the so-called \textit{monotonicity} of relative entropy under the inclusion of regions
\begin{equation}
S\left(\rho_1^A|\rho_0^A\right)\geq S\left(\rho_1^B|\rho_0^B\right)\,.\label{monotonicity2}
\end{equation}
Equation \reef{monotonicity2} basically tells us that if the states are already distinguishable when we compare them in a region $B\subseteq A$, they will be even more ``different'' when we contrast them in the larger region $A$.

The property of monotonicity \reef{monotonicity2} has been recently used to show that negative energy cannot be isolated far away from positive energy in a conformal field theory \cite{negative}. In a classical theory, the well known classical energy inequalities, that state the positivity of some combinations of the stress tensor components, are reasonable conditions postulated to hold in the theory (mainly aimed to prove theorems related to singularities). In particular, the so-called Weak Energy Condition (WEC) tell us for example that all observers measure positive values of energy density.

However, energy density in quantum field theory can take negative values if it is compensated by the presence of positive energy in other regions of space so as it is assured that the total energy is positive. In fact, in any QFT there are necessarily some states having negative energy density \cite{nega-necesaria}.

There have been various attempts in the past to quantify the amount of negative energy density allowed by quantum mechanics. In the literature, these sets of inequalities are referred to as \textit{quantum energy inequalities} (QEIs) \cite{QEI's}. While most of the inequalities found made statements about the duration in time of negative energy pulses, only several of them dealt with the constraints imposed to the spatial distribution of energy density. It is quite interesting that non-trivial inequalities of this type can be obtained using general properties of relative entropy.

For instance, the averaged null energy condition (ANEC), which is an important ingredient in the semi-classical proof of the generalised second law, has been recently proved to hold in general unitary and Lorentz invariant QFTs using the monotonicity property of relative entropy \cite{leigh}. A stronger inequality, the quantum null energy condition (QNEC), was later proved to hold too but its validity requires a more fine grained notion of causality, which is more than the monotonicity of relative entropy \cite{wang}.

The monotonicity property of relative entropy \reef{monotonicity2} has also proven to be useful in order to define interesting energy-entropy relations. The first inequality of this kind was due to Bekenstein \cite{bekenstein}, who derived an intriguing relation through a thought experiment involving black hole thermodynamics and classical physics. The validity and interpretation of this bound generated much discussion, until well defined forms were obtain from the property of positivity \cite{beke} and monotonicity \cite{negative} of relative entropy. This makes another fascinating case in which information theoretical tools provides us with valuable insights on subjects related to quantum field theory.
\\

\noindent \textbf{Outline}     \,\,\,In this work we use modular energy inequalities, derived from the property of monotonicity of the relative entropy \reef{monotonicity2}, in order to explore new QEIs and energy-entropy bounds. The paper is organised as follows. In section \ref{ch:modular} we explain how the properties of relative entropy can be used to produce inequalities for the expectation value of the modular hamiltonian. Modular hamiltonians are relevant objects that are sometimes related to the energy density; this is discussed in subsection \ref{ch:energy}. 
In section \ref{ch:QEIs} we use the modular energy relations in order to derive QEIs in a two-dimensional CFT. From these constraints, we arrive at some interesting conclusions related to the localization of negative energy in space, and find that the entropy of a state further restricts the possible amount of negative energy allowed by the theory. We show that a QEI derived from this procedure is in agreement and improves a previous bound by Fewster and Hollands \cite{fewster}.
In section \ref{ch:en-en}, we use the modular energy relations, to derive energy-entropy relations for CFTs.
We finish in section \ref{ch:discussion} with a review of the results obtained and we pose some questions that would be interesting to address in the future.

\section{Modular energy inequalities from relative entropy}\label{ch:modular}
%

Consider a state of a quantum field theory reduced to a region $V$, $\rho_V$.  Given that $\rho_V$ is a positive hermitian operator, it can always be written as
\begin{equation}
\rho_V=\frac{e^{-K_V}}{\tr\left(e^{-K_V}\right)}\,.
\end{equation}
$K_V$ is called the modular hamiltonian of the state $\rho_V$. It is simple to show that relative entropy between two states $\rho^1_V$ and $\rho^0_V$ reduced to a region $V$, can be written in terms of the modular hamiltonian $K_V$ corresponding to $\rho^0_V$ and the entanglement entropy of the states as \cite{first-law}
\begin{equation}
S\left(\rho^1_V | \rho^0_V\right)=\Delta \langle K_V \rangle - \Delta S_V\,,
\end{equation}
where $\Delta \langle K_V \rangle =  \langle K_V \rangle_1 - \langle K_V \rangle_0$ and $\Delta S_V = S_V^1 - S_V^0$. Therefore, if we consider two regions $A$ and $B$ such that $B \subseteq A$, from the property of monotonicity given by equation (\ref{monotonicity2}) we have
\begin{equation}
\Delta \langle K_A \rangle - \Delta S_A \geq \Delta \langle K_B \rangle - \Delta S_B \,,
\label{primera}
\end{equation}
and
\begin{equation}
\Delta \langle K_{\bar{B}} \rangle - \Delta S_{\bar{B}} \geq \Delta \langle K_{\bar{A}} \rangle - \Delta S_{\bar{A}} \,,
\label{segunda}
\end{equation}
since $\bar{A} \subseteq \bar{B}$.

Adding up equations (\ref{primera}) and (\ref{segunda}) we obtain the following inequality
\begin{equation}
\langle \hat{K}_A-\hat{K}_B \rangle _1 - \langle \hat{K}_A-\hat{K}_B \rangle _0 \geq \left(S_A^1-S_B^1+S_{\bar{B}}^1-S_{\bar{A}}^1 \right) - \left(S_A^0-S_B^0+S_{\bar{B}}^0-S_{\bar{A}}^0 \right)\,,
\label{desi}
\end{equation}
where $\hat{K}_X$ stands for the full modular hamiltonian of $X=A,\,B$ and it is defined as
\begin{equation}\label{eq:1}
\hat{K}_X = K_X-K_{\bar{X}}\,.
\end{equation}
Equation (20) in \cite{negative} is a particular case (when $\rho_0$ is the vacuum state) of the inequality (\ref{desi}) here.

Following \cite{negative}, we refer to the difference of entropies $S_A^1-S_B^1+S_{\bar{B}}^1-S_{\bar{A}}^1\equiv 2 S^1_F \left( A,B \right)$ as the \textit{free entropy} located in between the boundaries of $A$ and $B$. This free entropy $S_F$ is always positive as a consequence of the weak monotonicity property of entropy \cite{weak-mono} when applied to $A$ and $\bar{B}$
\begin{equation}
S_A+S_{\bar{B}} \geq S_{A-\bar{B}}+S_{\bar{B}-A} = S_B+S_{\bar{A}}\,.
\end{equation}
Interestingly, we can express the free entropy as $S_F(A,B)=\frac{I(A,\aleph)-I(B,\aleph)}{2}$, where $I(X,Y)=S(X)+S(Y)-S(X\cup Y)$ is the mutual information between $X$ and $Y$, and $\aleph$ is a hidden sector used to purify the state $\rho_1$. Since mutual information is a monotonically increasing quantity, the free entropy increases monotonically with the size $A-B$, though in general will not be an extensive quantity.

Using these definitions, inequality (\ref{desi}) becomes
\begin{equation}\label{eq:9}
\langle \hat{K}_A -\hat{K}_B \rangle_1 \ge 
  \langle \hat{K}_A -\hat{K}_B \rangle_0 -2S^0_F \left( A,B \right)+2S^1_F \left( A,B \right)\,,
\end{equation} 
where we must keep in mind that the all the modular hamiltonians involved are the ones corresponding to $\rho_0$.

A simpler inequality holds whenever $\rho_0$ is a pure state, i.e. $\rho_0=\vert \psi \rangle \langle \psi \vert$. Since the entanglement entropy of a pure state verifies $S_V=S_{\bar{V}}$ for any spatial region $V$, it is easy to show that $2S_F^0(A,B)=0$. On the other hand, the expectation value of $\ket{\psi}$ on the full modular hamiltonian (\ref{eq:1}) vanishes. This can be seen by expressing $\ket{\psi}$ in its Schmidt decomposition across the tensor product $\mathcal{H}_{\bar{X}} \otimes \mathcal{H}_{X}$ and writing the full modular hamiltonian in terms of its density operators $\hat{K}_X=\ln (\rho^0_{\bar{X}}) \otimes \mathbb{I}_X-\mathbb{I}_{\bar{X}}\otimes \ln (\rho^0_X)$. Therefore, the result is that, whenever $\rho_0$ is pure, inequality (\ref{eq:9}) becomes
\begin{equation}\label{eq:10}
\langle \hat{K}_A -\hat{K}_B \rangle_1 \ge 
  2S^1_f \left( A,B \right) \, .
\end{equation}
Notice that the information of the state $\ket{\psi}$ appears only through the modular hamiltonians in the left hand side.

These inequalities relate the ``modular energy'' (i.e. the expectation value of the modular hamiltonian) and the entropy in a non trivial manner. In some cases, the modular hamiltonian is related to the stress-energy tensor and as a consequence of this, the modular energy is related to the energy density of the state. We review this in detail in the following section.

\subsection{Comments on modular hamiltonians and its relation to energy}\label{ch:energy}

Modular hamiltonians are in general non-local objects and therefore the evolution they generate does not correspond to a local geometric flow. However, there are some remarkable cases in which the modular hamiltonian is explicitly known to be a local operator. For example, when we take $\rho_V$ as the vacuum state of any QFT reduced to the half spatial plane $V=\{x:x^0=0, x^1 > 0\}$, the modular hamiltonian asociated to $\rho_V$ is \cite{bisognano}
\begin{equation}
K_V=2\pi \int_{x^1 > 0} d^{d-1}x\, x^1\, T_{00}(x)\,.
\label{biso}
\end{equation}
This result follows from analicity properties originating in Lorentz invariance and positivity of energy. In this case, the modular hamiltonian is given by an integral of the energy density operator, weighted by the coordinate $x^1$ in which the region $V$ extends (this is simply the operator that generates the boost transformations in the plane $(x^0,x^1)$). Recently, the local modular hamiltonians of regions having its future horizon lying on a null plane were also found \cite{teste}.

Using equation (\ref{biso}) and conformal mappings, it is possible to obtain the modular hamiltonian of the vacuum state reduced to a ball of radius $R$ (we call this region $B$) for a CFT in $d+1$ dimensions \cite{sphere}
\begin{equation}\label{sphere}
K_B=2\pi \int_{B} d^{d}x\, \frac{R^2-x^2}{2R}\, T_{00}(x)\,.
\end{equation}
In the same way, the modular hamiltonian for the vacuum state of a CFT in a d-sphere $\mathbb{R}\times S^d$, reduced to a section $A$ of the sphere (given by $\phi \in [-\phi_A,\phi_A]$; $\phi$ is the azimuthal angle), is \cite{sphere,cylinder}
\begin{equation}\label{eq:cil}
K_A= 2\pi R \int d^{d-1}x\int\limits_{-\phi_A}^{\phi_A}d\phi\, \left(\frac{\cos (\phi)-\cos (\phi _A)}{\sin (\phi _A)}\right)T_{00}(\phi)\,.
\end{equation}

For two-dimensional CFTs, there are some other cases in which the modular hamiltonian of the vacuum is local and can be written again as an integral of the energy-momentum tensor times a local weight. A sufficient condition for this to happen is that the euclidean space-time region describing the traces of powers of the reduced density matrix (after removing small discs around the entangling points) is topologically an annulus \cite{tonni}.

In more general cases, the modular hamiltonian of the vacuum will naturally have non-local terms. Interestingly, for free massive scalar and fermionic fields in two-dimensional spacetime, the local part of the modular hamiltonian for any multi-interval region is also proportional to the stress tensor, with a universal coefficient independent of the mass that can be interpreted as a local temperature using relative entropy \cite{modular}.

For global states different from the vacuum state there are fewer results about the related modular hamiltonians. A remarkable result arises for a two-dimensional CFT in a thermal state at inverse temperature $\beta$ reduced to the half spatial line $V$. In this case, the modular Hamiltonian is a local object that can be expressed as an integral of the energy density \cite{yngvason}
\begin{equation}
K_V=\beta \int_{x > 0} dx \left( 1-e^{-2\pi x/\beta} \right) T_{00}(x)\,.
\label{thermal}
\end{equation}
An analogous expression holds when the region is an interval \cite{lashkari}.

We will use these results (particularly, equation (\ref{thermal})) to show how the modular energy inequalities (\ref{eq:9}) and (\ref{eq:10}) can be used to generate QEIs and energy-entropy bounds. 


\section{Quantum energy inequalities from modular energy relations}
\label{ch:QEIs}

In this section we derive QEIs from the modular energy relations (\ref{eq:9}) and (\ref{eq:10}), considering a two-dimensional CFT. We choose a particular theory, in which the symmetries allow us to obtain analytic expressions for each term in the modular energy inequalities.

\subsection{Quantum energy inequality from pure state}

We define the null coordinate $u_+=t+x$ and consider the ground state reduced to the region $A=\left\lbrace u_+  : u_+ \in (0,+\infty) \right\rbrace$. Its modular hamiltonian will be given by an expression equivalent to (\ref{biso}). From this, it is straightforward to get the full modular hamiltonian of $A$
\begin{equation}\label{eq:2}
\hat{K}_{A}^0=2\pi\int_{-\infty}^{+\infty}du_+ \, u_+ T_{++}(u_+)\, ,
\end{equation}
where $T_{++}(u_+)$ is the positive chiral component of the energy momentum tensor.\footnote{In a completely analogous way we can consider the problem in the other null direction $u_-=t-x$.} In order to apply a conformal transformation given by $u_+ \rightarrow u'_+=f(u_+)$, we use that the operator $T_{++}$ transforms according to the Schwartzian derivative $\left\lbrace f(u_+),u_+ \right\rbrace$ as
\begin{equation}\label{eq:7}
U^\dagger _f \, T_{++}(u_+) \, U_f=
  f'(u_+)^2T_{++}(f(u_+))-\frac{c}{24 \pi}\lbrace f(u_+),u_+ \rbrace\,,
\end{equation}
\begin{equation}\label{eq:6}
\lbrace f(u_+),u_+ \rbrace=
  \frac{f'''(u_+)}{f'(u_+)}-\frac{3}{2}\left(\frac{f''(u_+)}{f'(u_+)}\right)^2=
  -2\sqrt{f'(u_+)}\frac{d^2}{du_+^2}\frac{1}{\sqrt{f'(u_+)}}\, ,
\end{equation}
where $U_f$ is the unitary operator applying the conformal transformation. Using this in (\ref{eq:2}) we obtain the full modular hamiltonian of the transformed state $\ket{\psi} =U^\dagger _f \ket{0}$ reduced to the transformed region $A'=(f(0),f(+\infty))$
\begin{equation}
\hat{K}_{A'}^{\psi}=
  2\pi\int_{-\infty}^{+\infty}du_+ \, u_+ \left( f'(u_+)^2T_{++}(f(u_+))-\frac{c}{24 \pi}\lbrace f(u_+),u_+ \rbrace\right).
\end{equation}
For this derivation we recalled that the density matrix associated to $\hat{K}_{A'}^{\psi}$ is the operator which leads to the same expectation values on $\ket{\psi}$ for operators localised in $A'$ \footnote{A modular hamiltonian related to a global state $\ket{\psi}$ and region $V$ must verify 
\begin{equation}\label{eq:3}
\bra{\psi}\mathcal{O}_V\ket{\psi}=\tr \left[ e^{-K_V} \mathcal{O}_V\right],
\end{equation}
where this expression holds for every operator localized in the region of causal dependence of $V$. For instance, we may take $\mathcal{O}_V=\phi(x)$ with $x \in V$. We now want to find an analogous expression but for the transformed state $\ket{\Psi}=U^\dagger _f \ket{\psi}$. The left hand side of (\ref{eq:3}) can be written as

\begin{equation}
\bra{\psi}\left(U_f U_f^\dagger\right)\mathcal{O}_V \left(U_f U_f^\dagger\right)\ket{\psi}=
  \bra{\Psi}U_f^\dagger\mathcal{O}_V U_f \ket{\Psi}=\bra{\Psi}\mathcal{O}_{V'}\ket{\Psi}\,,
\end{equation}
where $V'$ is the transformed region. The right hand side of (\ref{eq:3}) can be written as 
\begin{equation}
\tr \left[e^{-K_V}\left(U_f U_f^\dagger\right) \mathcal{O}_V \left(U_f U_f^\dagger\right)\right]=
  \tr \left[U_f^\dagger e^{-K_V}U_f \mathcal{O}_{V'}  \right]=
  \tr \left[ e^{-U_f^\dagger K_V U_f} \mathcal{O}_{V'}  \right]\,.
\end{equation}

We therefore find that the modular hamiltonian transforms under the conformal transformation as
\begin{equation}
K_V^{\psi} \longrightarrow K_{V'}^{\Psi}=U_f^\dagger K_V^{\psi} U_f\,.
\end{equation}
}. In a completely analogous way we can write the same full modular hamiltonian but reduced to the region $B'= (f(a),f(+\infty))$ with $a$ a positive constant. Hence, we find

\begin{equation}\label{eq:4}
\hat{K}_{A'}^{\psi}-\hat{K}_{B'}^{\psi}=2\pi
  a\int\limits_{-\infty}^{+\infty}du_+\, \,
  \left(
  f'(u_+)^2T_{++}(f(u_+))-\frac{c}{24 \pi}\lbrace f(u_+),u_+ \rbrace
  \right).
\end{equation}

Considering this expression in the modular energy inequality (\ref{eq:10}) with $\rho_0=\ket{\psi}\bra{\psi}$ we find 
\begin{equation}\label{eq:5}
\int\limits_{-\infty}^{+\infty}du_+\, \,  
  f'(u_+)^2 \langle T_{++}(f(u_+))\rangle _1
 \ge \frac{c}{24 \pi} \int\limits_{-\infty}^{+\infty}du_+\,\lbrace f(u_+),u_+ \rbrace
 +\frac{1}{\pi a}S_F^1(A',B')\,.
\end{equation}

This inequality is already a QEI valid for \textit{any} state $\rho_1$ in a two-dimensional CFT. We recognise the similarity of this inequality with a previous one derived in \cite{fewster}. In fact, if we use the explicit form of the Schwartzian derivative (\ref{eq:6}), integrate by parts the right hand side and change the integration variable to $f(u_+)$ we get
\begin{equation}
\int\limits_{-\infty}^{+\infty}du_+\, \,  
  g(u_+) \langle T_{++}(u_+)\rangle _1
 \ge -\frac{c}{12 \pi} \int\limits_{-\infty}^{+\infty}du_+\,\left(\frac{d}{du_+}\sqrt{g(u_+)}\right)^2
 +\frac{1}{\pi a}S_F^1(A',B'),
\label{eq:8}
\end{equation}
where we defined $g(u_+)=f'(f^{-1}(u_+))$ (and we assumed that $f$ is a diffeomorphism of $\mathbb{R}$ with $f\left(+\infty\right)=+\infty$ and $f''/f'\rightarrow 0$ for $x\rightarrow \pm\infty$).

Equation (\ref{eq:8}) is a stronger version of an inequality without the free entropy term, that arises from (\ref{eq:7}) and the positivity of energy \cite{fewster}, namely

\begin{equation}
\int\limits_{-\infty}^{+\infty}du_+\, \,  
  g(u_+) \langle T_{++}(u_+)\rangle _1
 \ge -\frac{c}{12 \pi} \int\limits_{-\infty}^{+\infty}du_+\,\left(\frac{d}{du_+}\sqrt{g(u_+)}\right)^2\,.
 \label{few}
\end{equation}

This last inequality was proven to hold for every function $g(u_+)$ of the Schwartz class, with the rhs being the infimum of the lhs as the state $\rho_1$ varies within a certain dense subspace of the Hilbert space. So, if $\rho_c$ is the optimal state for which inequality (\ref{few}) saturates, our strongest version of the inequality, equation (\ref{eq:8}) tell us that the free entropy of $\rho_c$ over the regions $A'$ and $B'$ must vanish (this is indeed the case whenever the optimal state $\rho_c$ is pure, regardless of the regions $A'$ and $B'$). The saturation of equation (\ref{eq:8}) can be related to the saturation of the relative entropies

\begin{equation}
S(\rho_c^{A'}|\rho_0^{A'})=S(\rho_c^{B'}|\rho_0^{B'})\,,
\label{satu1}
\end{equation}
and

\begin{equation}
S(\rho_c^{\bar{A}'}|\rho_0^{\bar{A}'})=S(\rho_c^{\bar{B}'}|\rho_0^{\bar{B}'})\,,
\label{satu2}
\end{equation}
where $\rho_0=U^\dagger _f \ket{0}\bra{0}U_f$, $A'=(f(0),f(+\infty))$ and $B'=(f(a),f(+\infty))$.

In the derivation of \cite{fewster}, the pure state that saturates the QEI is $\rho_c=\rho_0$; this is related to a trivial saturation of the monotonicity property (\ref{monotonicity}), since each relative entropy in equations (\ref{satu1}) and (\ref{satu2}) is zero. In general, it is expected that other states may accomplish the task of saturating inequality (\ref{monotonicity2}), for non-zero relative entropies in the relations (\ref{satu1}) and (\ref{satu2}). The saturation of the monotonicity of relative entropy is an interesting mathematical problem that has been discussed in the literature \cite{monot1}, and the result we have obtained might be useful in the understanding of it.

It is clear though that for general states (non necessarily pure states), if $\rho_1 \neq \rho_c$, the free entropy term in equation (\ref{eq:8}) improves the bound given by (\ref{few}). In the following section we show how to use this new bound to obtain a new quantum energy inequality.

\subsection{Quantum energy inequality from mixed state}\label{ch:ejemplo}

\begin{figure}[h!]
\begin{center}
\includegraphics[width=8cm]{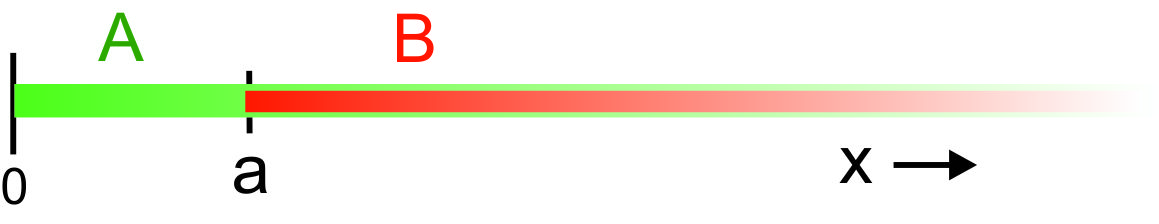}
\end{center}
\caption{\textbf{Diagram of the spatial regions $A$ and $B$ considered.} Both $A$ and $B$ are half spaces, chosen so that $B \subseteq A$.} \label{Fig0}
\end{figure}

We now consider inequality (\ref{eq:9}), taking $\rho_0$ as a Gibbs thermal state with temperature $T=1/\beta$. We take the regions $A$ and $B$ as the half spaces given by $x\geq 0$ and $x\geq a$ respectively, with $a > 0$ (Figure \ref{Fig0}) so that $B \subseteq A$.

The modular hamiltonian of the thermal state reduced to these regions and its complements can be read from equation \reef{thermal}. This gives us the full modular hamiltonians for regions $A$ and $B$
\begin{equation}
\hat{K}_A = \int_{-\infty}^{+\infty} dx\,  \beta \left(1-e^{-2\pi \vert x \vert /\beta}\right) sgn\left(x\right)T_{00}\left(x\right)\,,
\end{equation}
and
\begin{equation}
\hat{K}_B = \int_{-\infty}^{+\infty} dx\,  \beta \left(1-e^{-2\pi \vert x-a \vert /\beta}\right) sgn\left(x-a\right)T_{00}\left(x\right)\,;
\end{equation}
its difference may be cast as
\begin{equation}
\hat{K}_A -\hat{K}_B = \int_{-\infty}^{+\infty} dx\, \beta f\left(x\right) T_{00}\left(x\right) \,,\label{difejam}
\end{equation}
where we have defined $f\left(x\right)= \left(1-e^{-2\pi \vert x \vert /\beta}\right) sgn\left(x\right)-\left(1-e^{-2\pi \vert x-a \vert /\beta}\right) sgn\left(x-a\right)$. Consequently, the left hand side (lhs) of equation (\ref{eq:9}) is
\begin{equation}
\langle\hat{K}_A -\hat{K}_B \rangle _1 = \int_{-\infty}^{+\infty} dx\, \beta f\left(x\right) \langle T_{00}\left(x\right) \rangle _1 \,.
\end{equation}
The only information that remains in the last expression about the particular state $\rho_0$ is in the function $f$ that weights the energy density in the integral.\footnote{A different choice of $\rho_0$ would change not only the function $f(x)$ but might as well change the operator that appears in the expression and even the form of the expression (in fact, for an arbitrary $\rho_0$ the modular hamiltonian will not be a local operator in general).}

We can now move to analyse the rhs of equation (\ref{eq:10}) and evaluate $\langle \hat{K}_A-\hat{K}_B \rangle _0$ for the thermal state $\rho^0$. This is straightforward, since for thermal states the energy density is constant and equal to $\langle T_{00}\left(x\right) \rangle _0 = \frac{c}{6}\frac{\pi}{\beta^2}$ \cite{difrancesco}. Then
\begin{equation}\label{eq:11}
\langle \hat{K}_A-\hat{K}_B \rangle _0 = \frac{c}{3}\frac{\pi a}{\beta}\,.
\end{equation}

In order to calculate the free entropy of $\rho_0$ in (\ref{eq:10}), we make use of the result for the entropy of a thermal state reduced to an interval of lenght $\Lambda$ in a two-dimensional CFT \cite{cacardy}
\begin{equation}
S\left(\Lambda\right)=\frac{c}{3}\log \left(\frac{\beta}{\pi\epsilon}\sinh\left(\frac{\pi\Lambda}{\beta}\right)\right)\,.
\label{entrotermo}
\end{equation}
$\epsilon$ is an ultraviolet cutoff used to regulate the divergences that come from short-distance entanglement around the border of the region and $c$ is the central charge of the Virasoro algebra. Since we need the entropies for states reduced to half space, we can use equation (\ref{entrotermo}) regarding $\Lambda$ as an infrared regulator. For the difference of entropies appearing in the free entropy, the limit $\Lambda \rightarrow +\infty$ gives us a finite, UV and IR regularisation-independent result
\begin{equation}\label{eq:12}
2S_F^0(A,B)=S_A^0-S_B^0+S_{\bar{B}}^0-S_{\bar{A}}^0=\frac{2c}{3}\frac{\pi a}{\beta}\,.
\end{equation}
Inserting the results of (\ref{difejam}), (\ref{eq:11}) and (\ref{eq:12}) into equation (\ref{eq:10}), we obtain a new quantum energy inequality valid for \textit{any} state $\rho_1$ of a two-dimensional CFT and arbitrary positive constants $a$ and $\beta$
\begin{equation}
\int_{-\infty}^{+\infty} dx\, f\left(x\right) \langle T_{00}\left(x\right)\rangle _1 \geq - \frac{c}{3}\frac{\pi a}{\beta^2}+\frac{2}{\beta}S_F^1(A,B)\,,\label{cotita}
\end{equation}
with
\begin{equation}\label{eq:14}
f\left(x\right)=\left(1-e^{-2\pi \vert x \vert /\beta}\right) sgn\left(x\right)-\left(1-e^{-2\pi \vert x-a \vert /\beta}\right) sgn\left(x-a\right)\,.
\end{equation}
As we will show next, this equation imposes severe constraints to the distributions of energy for the states of the theory.

\subsection*{Analysis of the constraint}

The entropy contribution of the QEI (\ref{cotita}) cannot be calculated without specifying the state $\rho_1$. However, since it is non-negative, we can analyse a weaker constraint that does not consider its contribution, i.e.

\begin{equation}
\int_{-\infty}^{+\infty} dx\, f\left(x\right) \langle T_{00}\left(x\right)\rangle _1 \geq - \frac{c}{3}\frac{\pi a}{\beta^2}\,.
\end{equation}

First, notice that $f(x)$ is a dimensionless function that depends only on the parameters $a$ and $\beta$. For a fixed value of $a/\beta$, the graphic of the function has the same form but its typical size varies for different values of $a$ (notice that $\int_{-\infty}^{+\infty}dx f\left(x\right)=2a$, for all values of $a$ and $\beta$).

\begin{figure}[h]
\begin{center}
\includegraphics[width=10cm]{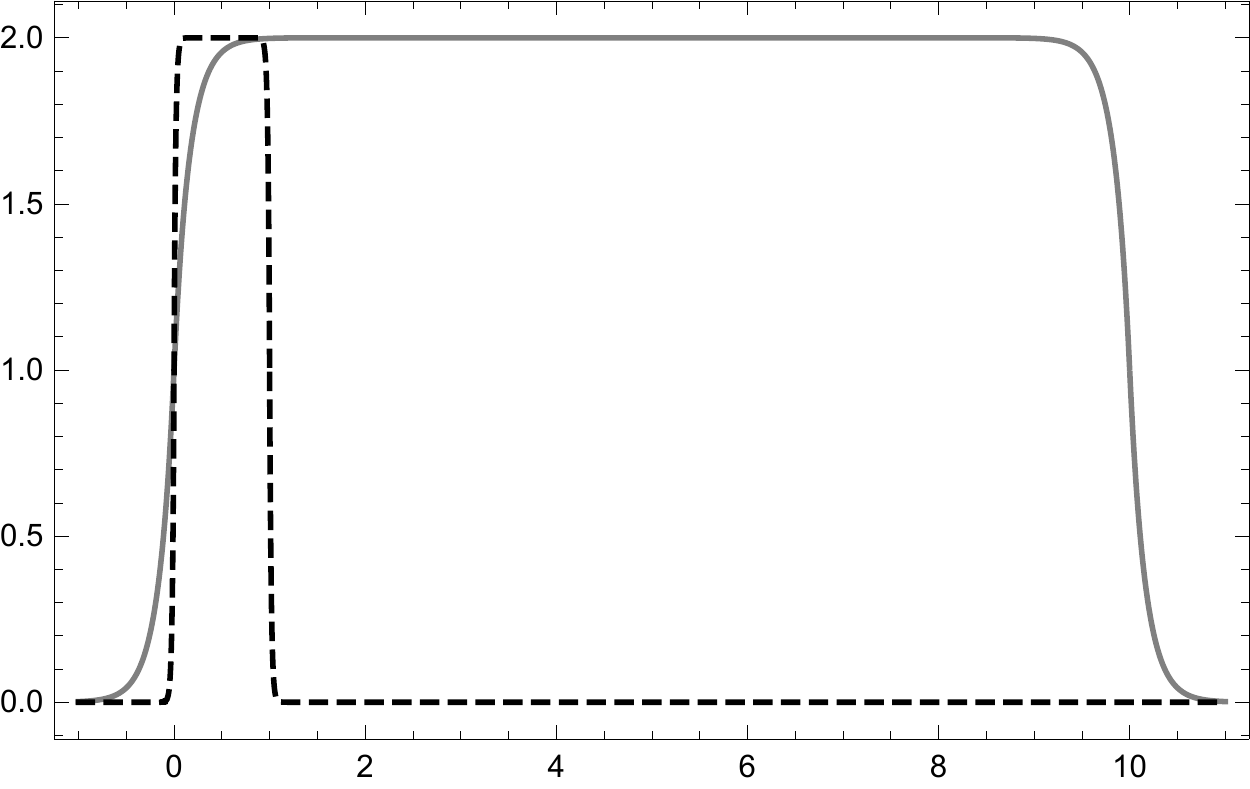}
\end{center}
\caption{\textbf{Analysis of $f(x)$ for large values of $a/\beta$.} For a fixed large value of $a/\beta$ (in this case, $a/\beta=10$) the graph of the function $f$ tends to the square barrier $\textrm{sgn}(x)-\textrm{sgn}(x-a)$ and has a sharp slope at the points $x=0$ and $x=a$. Dotted line corresponds to $a=1$, full line to $a=10$.} \label{ventana1}
\end{figure}

To see this behaviour, in figure \ref{ventana1} we plot the function $f(x)$ for a fixed value $a/\beta=10$ with $a=10,\,1$ (dotted line corresponds to $a=1$; notice that $a$ is basically the support of the function, when $a/\beta$ is large). In this limit, for the three chosen values of $a$, the graphic of the function $f(x)$ is approximately a square-like barrier given by $\textrm{sgn}(x)-\textrm{sgn}(x-a)$.

\begin{figure}[h]
\begin{center}
\includegraphics[width=10cm]{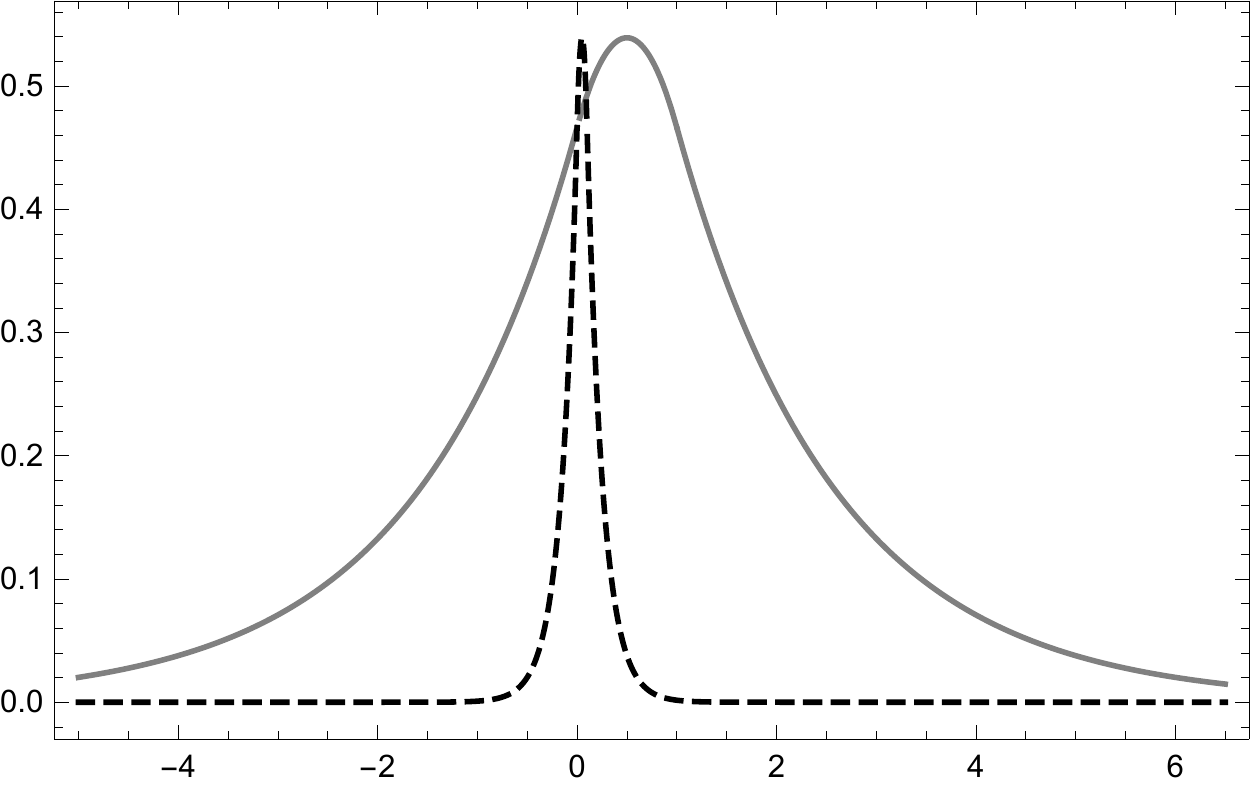}
\end{center}
\caption{\textbf{Analysis of $f(x)$ for small values of $a/\beta$.} For a fixed small value of $a/\beta$ (in this case, $a/\beta=1/10$) the graph of the function $f$ is a smooth symmetric bell centered at $x=a/2$. As we increase the value of $a$, keeping $a/\beta$ fixed, the bell spreads more in space. Dotted line corresponds to $a=0.1$, full line to $a=1$.} \label{ventana2}
\end{figure}

As we take smaller values of $a/\beta$, the function gets smoother at the points $x=0$ and $x=a$. The graph of the function (see figure \ref{ventana2}) is a symmetric bell centered at $x=a/2$ that spreads more in space as we take bigger values of $a$, while keeping $a/\beta$ fixed.

\begin{figure}[h]
\begin{center}
\includegraphics[width=12cm]{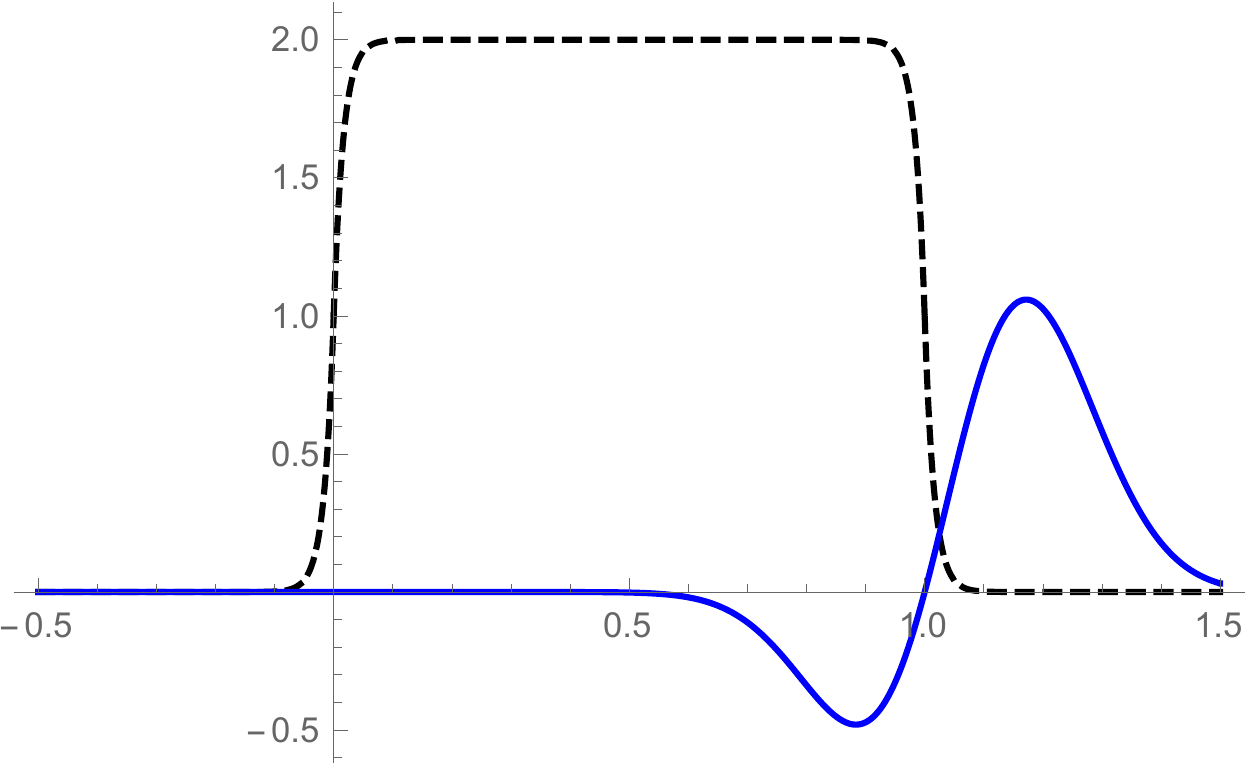}
\end{center}
\caption{\textbf{Analysis of the QEI for large values of $a/\beta$.} The energy density (solid blue line) consists of a negative and positive pulse, with the positive one being suppressed by the function $f$ (dashed black line). This tell us that the lhs \reef{cotita} will take large negative values and this is in agreement with the large negative values for the rhs of \reef{cotita}, given when $a/\beta$ is large.} \label{pulsos}
\end{figure}

We can explore the implications of equation \reef{cotita} in the limit $a/\beta \gg 1$. In this case, we have seen that the function tends to a square barrier (see figure \ref{ventana1}). On the other hand, the right hand side (rhs) of equation \reef{cotita} is a large negative number in this limit. This can be understood as follows. Suppose that we can construct a state $\rho_1$ whose energy density consists of a pulse of negative energy and another pulse of positive energy distributed in space as sketched in figure \ref{pulsos} (solid blue line). The function $f$ in this limit is also plotted in figure \ref{pulsos} (dashed black line). Notice that the total energy of the state will be a positive number, as required. However, since the positive pulse is mainly localised outside the region $0<x<a$, while the negative one is mostly inside, the left hand side of equation \reef{cotita} will effectively take a large negative value. This is in harmony with the increasing negative values for the right hand side of equation \reef{cotita} when $a/\beta$ is a large number.

\begin{figure}[h]
\begin{center}
\includegraphics[width=12cm]{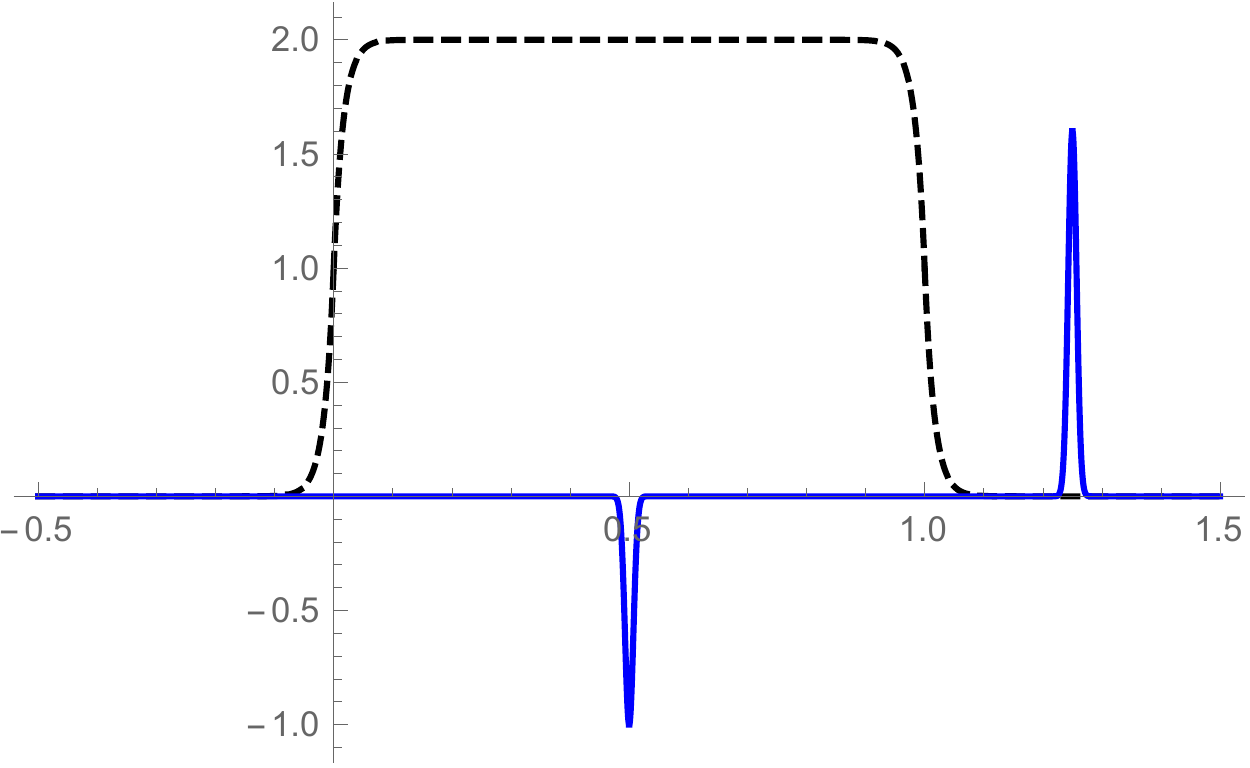}
\end{center}
\caption{\textbf{Constraints to the negative energy allowed in a CFT.} In this situation, the lhs of equation \reef{cotita} will be twice the total negative energy. For the inequality \reef{cotita} to hold, it is necessary that the total negative energy of the pulse is bounded.} \label{pulsos2}
\end{figure}

Notice though, that for a fixed $f(x)$ (taken so as $a/\beta$ is large) the right hand side of equation \reef{cotita} will be fixed. This imposes a restriction to the allowed physical states of the theory. For instance, we can arbitrarily increase the magnitude of both the negative and positive energy pulses shown in figure \ref{pulsos2} while keeping $E=E_+-E_-\geq 0$ ($E_\pm=\int dx\,\,\theta(\pm\langle T^{00}(x)\rangle)\,\,|\langle T^{00}(x)\rangle_1|$), but equation \reef{cotita} will be violated in this case if the total negative energy $E_-$ is, roughly speaking, larger than $\dfrac{c\pi a}{6\beta^2}$. $a$ and $\beta$ are arbitrary, and for the last reasoning to apply they only need to satisfy that $a/\beta \gg 1$. This can be accomplished, for example, by taking $a=d$ (the distance between the negative and positive energy pulses) and $\beta=r_-$ (the dispersion of the negative energy density). In the situation represented on figure \ref{pulsos2}, the analysis we made is valid and we can also see that $a/\beta=d/r_-$ can be as large as we want by moving the positive pulse away from the negative one (and keeping its dispersion $r_+$ small enough so that the positive part of the energy density does not fall under the bell of figure \ref{pulsos2} \footnote{Alternatively, one could have taken $a=r_+$ and make it large enough so that $a/\beta=r_+/r_- \gg 1$ by separating the pulses long enough (in order that the positive energy pulse falls outside of the bell of figure \ref{pulsos2}). This would give us $E_- r_-^2\leq \frac{c\pi r_+}{6}$, instead of equation \reef{cota}; in the situation that we set both equations have the same implications.}). Therefore, we arrive at
\begin{equation}
E_- r_-^2\leq \frac{c\pi d}{6}\,.
\label{cota}
\end{equation}
This last equation tell us for example that we can increase the total amount of negative energy at expense of reducing its dispersion. This relation, where the intrinsic size of the negative energy ``moment of inertia'' is bounded from above by moments of the positive energy distribution, is similar to the one found in \cite{negative}.

\subsection{Comparison of the results}

In this section we compare the QEI derived from a pure state (\ref{eq:8}) with the one derived using a thermal state (\ref{cotita}). Both inequalities have a different contribution coming from the entropy, since the regions considered differ. However, since the free entropy is non-negative, we can compare the weaker bounds that do not consider these contributions
\begin{equation}
\int\limits_{-\infty}^{+\infty}dx\, \,  
  g(x) \langle T_{00}(x)\rangle _1
 \ge -\frac{c}{6 \pi} \int\limits_{-\infty}^{+\infty}dx\,\left(\frac{d}{dx}\sqrt{g(x)}\right)^2\,,
\label{eq:13}
\end{equation}
\begin{equation}
\int\limits_{-\infty}^{+\infty} dx\, f\left(x\right) \langle T_{00}\left(x\right)\rangle _1 \geq - \frac{c}{3}\frac{\pi a}{\beta^2}\,,
\label{eq:15}
\end{equation}
where we have used that $T_{00}(x)=T_{++}(x)+T_{--}(-x)$ and the function $f(x)$ is given by (\ref{eq:14}). 

In order to compare \reef{eq:13} and \reef{eq:15}, we might simply consider (\ref{eq:13}) with $g(x)=f(x)$. We have to note though, that Fewster and Hollands showed that the function $g(x)$ must belong to the Schwartz class, and the function $f(x)$ does not fulfill this requirement. Nonetheless, we can calculate the integral from the rhs of \reef{eq:13} for a set of Schwartzian functions $\lbrace f_n(x) \rbrace$ such that $f_n(x) \rightarrow f(x)$ when $n\rightarrow \infty$, and then take the limit of the succession obtained\footnote{We have done this using two sets of approximating functions for $|x|$ and $\textrm{sgn}(x)$ (i.e. $m^{(1)}_n(x)=\sqrt{x^2+1/n^2}$ and $m^{(2)}_n(x)=\frac{x^2}{\sqrt{x^2+1/n^2}}$ for $|x|$, and $s^{(1)}_n(x)=\frac{x}{\sqrt{x^2+1/n^2}}$ and $s^{(2)}_n(x)=\tanh\left(nx\right)$ for $\textrm{sgn}(x)$) and in all the cases we get the same result for the rhs integral of equation \reef{eq:13} in the limit $n\rightarrow \infty$.}. This procedure is reasonable, since $f(x)$ can be approximated as much as we want by a set of Schwartian functions and therefore, for any physically reasonable distribution of energy density, the result will be insensible to the arbitrarily small differences between $f(x)$ and $f_n(x)$ for large $n$ (notice also that in order to compute the rhs of equation \reef{eq:13} it is sufficient for the function to have one derivative - our function $f(x)$ in (\ref{eq:14}) is not of the Schwartian type but has a well-defined first derivative).

With this in mind, the result given by (\ref{eq:13}) is
\begin{equation}\label{eq:16}
\int\limits_{-\infty}^{+\infty} dx\, f\left(x\right) \langle T_{00}\left(x\right)\rangle _1  \geq \frac{c}{3} \frac{\pi a}{\beta^2}-\frac{c}{3 \beta} \sqrt{1-e^{-2\pi a/\beta}} \textrm{arctanh}\left(\sqrt{1-e^{-2\pi a/\beta}}\right)\,.
\end{equation}

\begin{figure}[h!]
\begin{center}
\includegraphics[width=10cm]{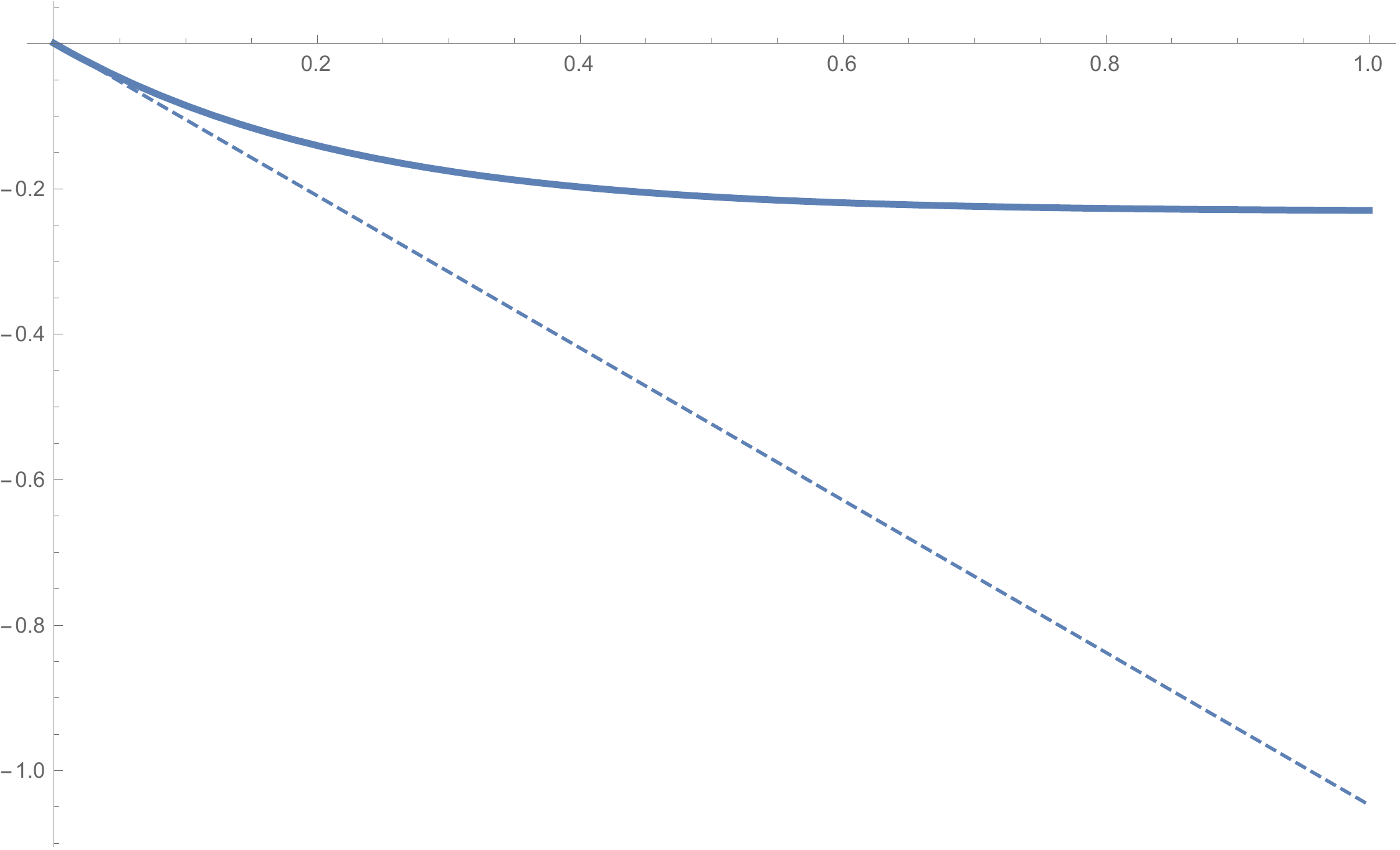}
\end{center}
\caption{\textbf{Comparison of the lower bounds for the expectation value of the energy density weighted by the function f, without the entropy terms.} The straight dashed line represents the rhs of equation \reef{eq:13} while the solid line is the rhs of equation \reef{eq:15}. The inequalities are compatible with the previous result by Fewster and Hollands, and the two bounds happen to coincide at the lowest order for small $a/\beta$. It is important to recall that we are not considering the free entropy terms of the QEIs.} \label{pucha}
\end{figure}

To see how this compares with inequality (\ref{eq:15}), in figure \ref{pucha} we plot the obtained lower bounds for the expectation value of the energy density weighted by the function $f(x)$ as a function of the parameter $a$, i.e., the rhs of equations \reef{eq:13} and \reef{eq:15} (we take $c=1$ and $\beta=1$). The straight dashed line represents the lower bound obtained from the thermal state, while the solid line from considering a pure state. We can see that the results are compatible. In fact, they coincide at first order in $a/\beta$ as can be easily seen by expanding the rhs of \reef{eq:16} in powers of $a/\beta$.

It is not surprising that (\ref{eq:13}) is more restrictive, since we know that the bound is sharp in that case.

\section{Energy-entropy bounds from modular energy relations}
\label{ch:en-en}

In this section, we derive interesting energy-entropy relations, analogous to the Bekenstein bound. These inequalities arise from the energy relations (\ref{eq:9}) and (\ref{eq:10}).

\subsection{Two-dimensional CFT}

First, we derive an energy-entropy bound from the QEI previously obtained in section \ref{ch:ejemplo}. Consider a state $\rho_1$ that has a localised energy density, i.e., $\left\langle T_{00}(x) \right\rangle =0$ outside the region $\left(0,L\right)$ and take $a=L$. In the limit of $L/\beta \ll 1$, inequality (\ref{eq:15}) becomes
\begin{equation}
2\pi L E \geq -\frac{c \pi L}{3\beta} + S_F^1\left(A,B\right)\,,
\end{equation}
where $E$ is the energy of $\rho_1$. Since $L/\beta \ll 1$, we have
\begin{equation}
S_F^1\left(A,B\right) \ll 2\pi E L\,. \label{bekelike}
\end{equation}
A similiar inequality was previously found on \cite{negative}, but for a CFT in any number of dimensions. Both results turn up to be consistent. The fact that the free entropy is bounded linearly in $L$ is consistent with the monotonic behaviour of $S_F^1(A,B)$ with $A-B$.

\subsection{d+1-dimensional CFT}

\begin{figure}[h]
\begin{center}
\includegraphics[width=9cm]{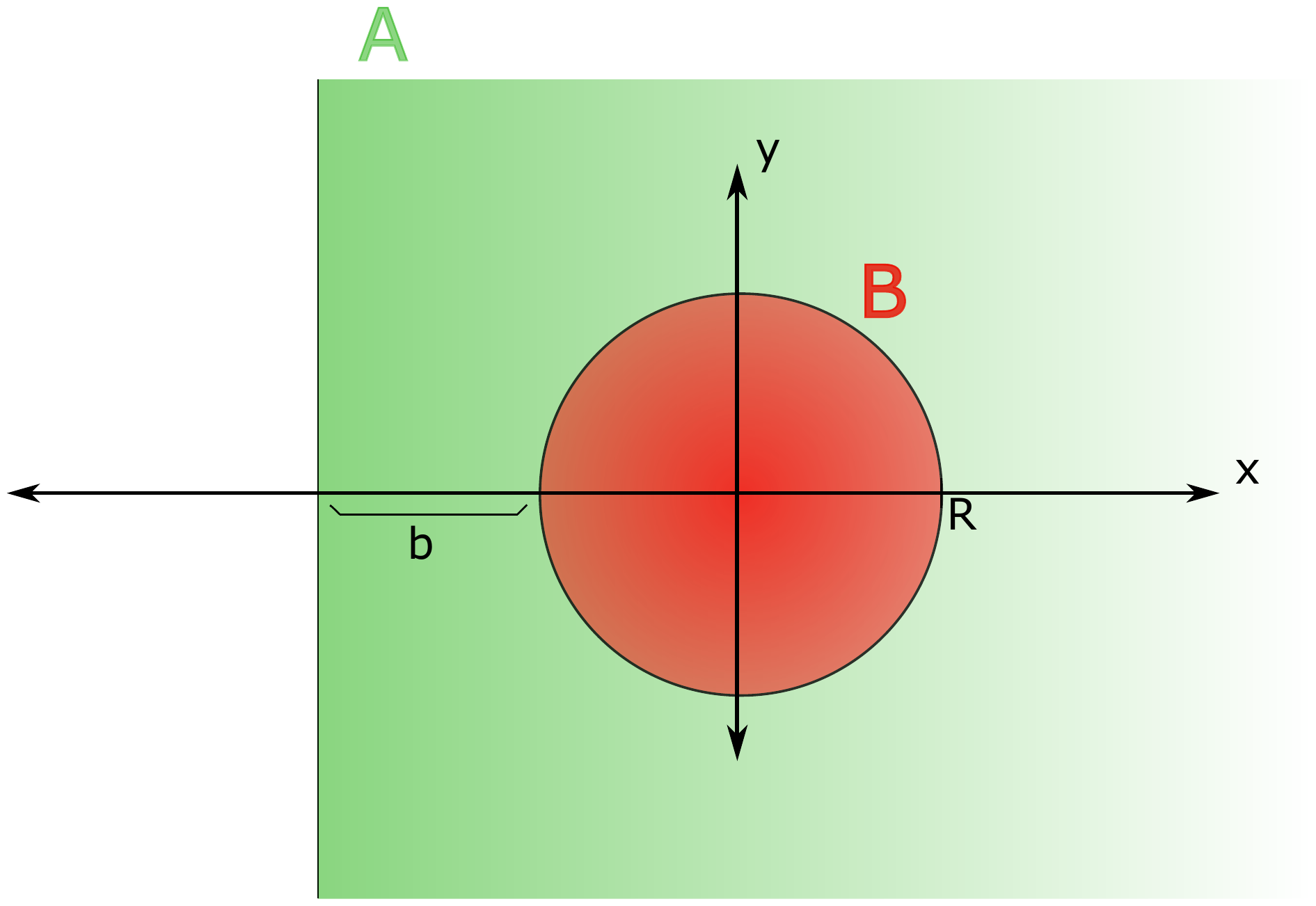}
\end{center}
\caption{\textbf{Diagram of the spatial regions $A$ and $B$ considered for a fixed time and two spatial dimensions.} The parameter $b$ must verify $b \geq 0$ so that $B \subseteq A$.} \label{regions1}
\end{figure}

We now consider a CFT in $\mathbb{R}\times \mathbb{R}^{d}$ in its ground state, take the region $B$ as a sphere of radius $R$ and the region $A$ as the half the space beginning at a distance $R+b$ from the origin of the sphere with $b \ge 0$ so that we have $B \subseteq A$. Figure \ref{regions1} shows a diagram of the spatial regions considered when $d=2$.

The modular hamiltonians of the ground state reduced to the regions $A$ and $B$ can be read from (\ref{biso}) and (\ref{sphere}), so that the full modular hamiltonians are equal to
\begin{equation}
\hat{K}_B=2 \pi \int d^{d}x\,\frac{R^2-|\vec{x}|^2}{2R}T_{00}(\vec{x})\,,
\end{equation}
and
\begin{equation}
\hat{K}_A=2\pi \int d^dx\,(x^1+R+b) T_{00}(\vec{x})\,,
\end{equation}
where both integrals are over the whole space. Since the global state is the vacuum which is pure, we use the modular energy inequality (\ref{eq:10}) and find 
\begin{equation}\label{eq:12.}
S^1_f(A,B) \le  \frac{\pi}{2R} \int d^dx \Big ( |\vec{x}|^2+2Rx^1+R(R+2b)\Big ) \langle T_{00}(\vec{x}) \rangle _1 ,
\end{equation}
which holds for every positive value of $R$ and $b$. This inequality can be written in a way that is more enlightening, by defining the ``center of energy'' $\vec{x}_e$
\begin{equation}
\vec{x}_e=\frac{1}{E}\int d^dx\,\vec{x}\langle T_{00}(\vec{x}) \rangle\,,
\end{equation}
and its dispersion $r_e$
\begin{equation}
r^2_e=\frac{1}{E}\int d^dx\,|\vec{x}-\vec{x}_e|^2\langle T_{00}(\vec{x}) \rangle \,,
\end{equation}
so that (\ref{eq:12.}) becomes
\begin{equation}
S^1_f(A,B) \le   \frac{\pi }{2} E  \left[ \frac{|\vec{x}_e|^2+r^2_e}{R}+R+2(b+x^1_e) \right]\,.
\end{equation}
This expression is already an energy-entropy inequality that holds for every positive value of $R$ and $b$. We use the freedom to choose the center of coordinates in order to minimize the rhs of the last equation (this can be achieved by taking $x^i_e=-R\delta^{i1}$). This results in
\begin{equation}\label{eq:17}
S^1_f(A,B) \le   \frac{\pi }{2} E  \left(\frac{r^2_e}{R}+2b \right)\, . 
\end{equation}
The bound holds for any state $\rho_1$ in the CFT and is consistent with the fact that the free entropy increases monotonically with the size of the region $A-B$, since as $b$ increases and $R$ decreases, $A-B$ grows.

\subsection{CFT in a cylinder}

\begin{figure}[h]
\begin{center}
\includegraphics[width=5cm]{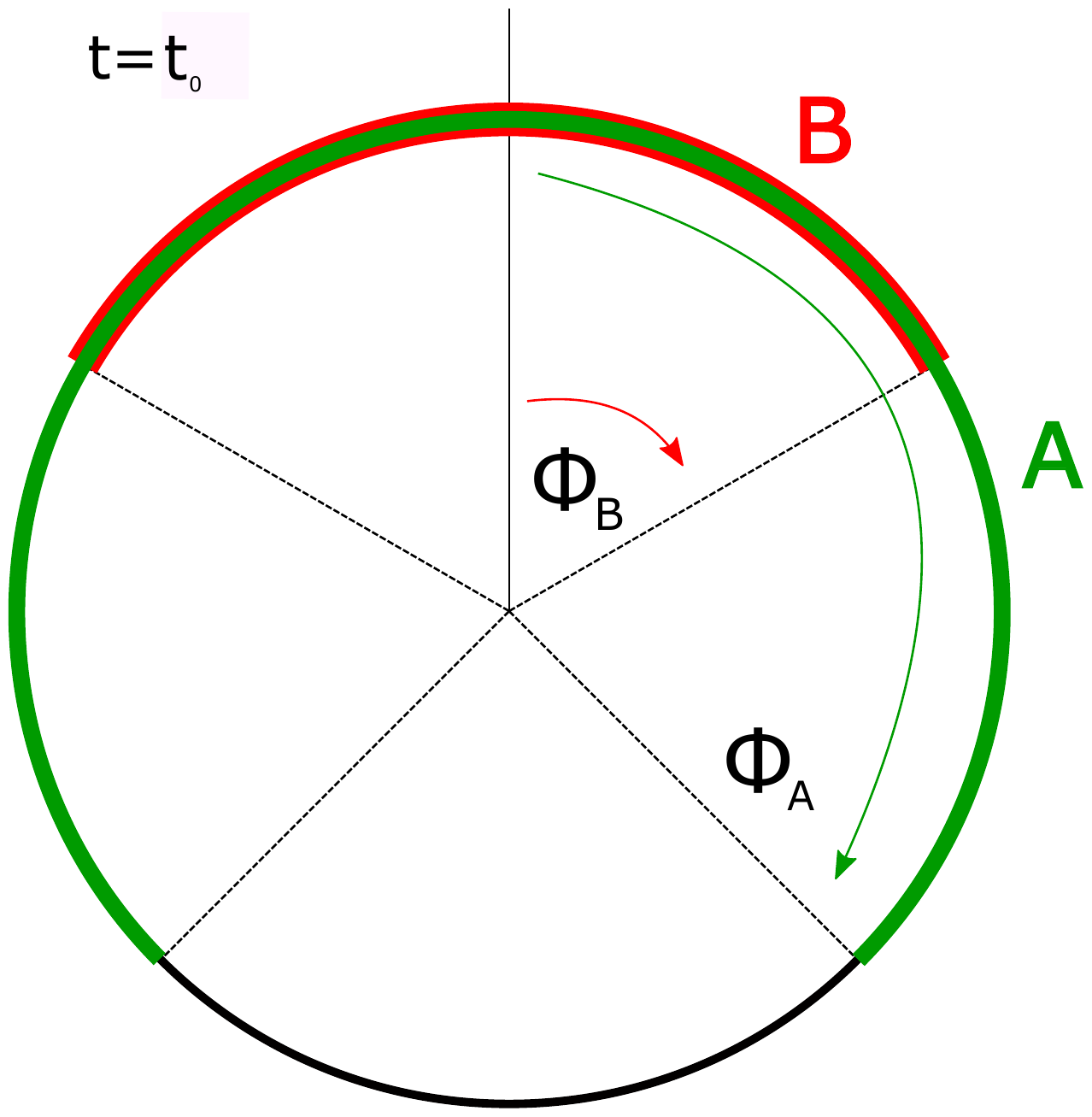}
\end{center}
\caption{\textbf{Diagram of the spatial regions $A$ and $B$ considered for $d=1$.} $\phi$ is the azimuthal angle of the $d$ sphere. The parameters $\phi_A$ and $\phi_B$ must verify $0 \le \phi_B \le \phi _A \le \pi$ so that the condition $B  \subseteq A$ holds.} \label{regions2}
\end{figure}

We take a CFT in a cylindrical space-time $\mathbb{R}\times S^d$ of radius $R$ in its ground state. We define the spatial regions $A$ and $B$ as sections of the sphere determined by $[-\phi _A , \phi _A]$ and $[-\phi _B , \phi _B]$, where $0 \le \phi_B \le \phi _A \le \pi$ so that $B  \subseteq A$. The angle $\phi$ is the azimuthal angle $\phi \in [0,2\pi)$ of the $d$ sphere. Figure \ref{regions2} shows a diagram of the spatial regions considered when $d=1$.

The modular hamiltonian of the ground state reduced to $A$ is given by (\ref{eq:cil}). From this expression it is straightforward to write the full modular hamiltonian of $A$ as
\begin{equation}
\hat{K}_A= 2\pi R \int d^{d}x\,\left(\frac{\cos (\phi)-\cos (\phi _A)}{\sin (\phi _A)}\right) T_{00}(\phi)\,,
\end{equation}
where the integral is over the whole space. In an analogous way we can write $\hat{K}_B$. After some algebra, the modular energy inequality (\ref{eq:10}) becomes
\begin{equation}\label{eq:4.}
S_F^1(A,B) \le \pi R \int d^{d}x \,\left(\frac{\sin (\phi _A- \phi _B)+\cos (\phi)\left[\sin(\phi _B)-\sin (\phi _A)\right]}{\sin (\phi _A)\sin (\phi _B)}\right)\langle T_{00} (\phi)\rangle_1,
\end{equation}
where we must have $0 \le \phi_B \le \phi _A \le \pi$. Equation \reef{eq:4.} is another energy-entropy inequality.

\begin{figure}[h]
\begin{center}
\includegraphics[width=6cm]{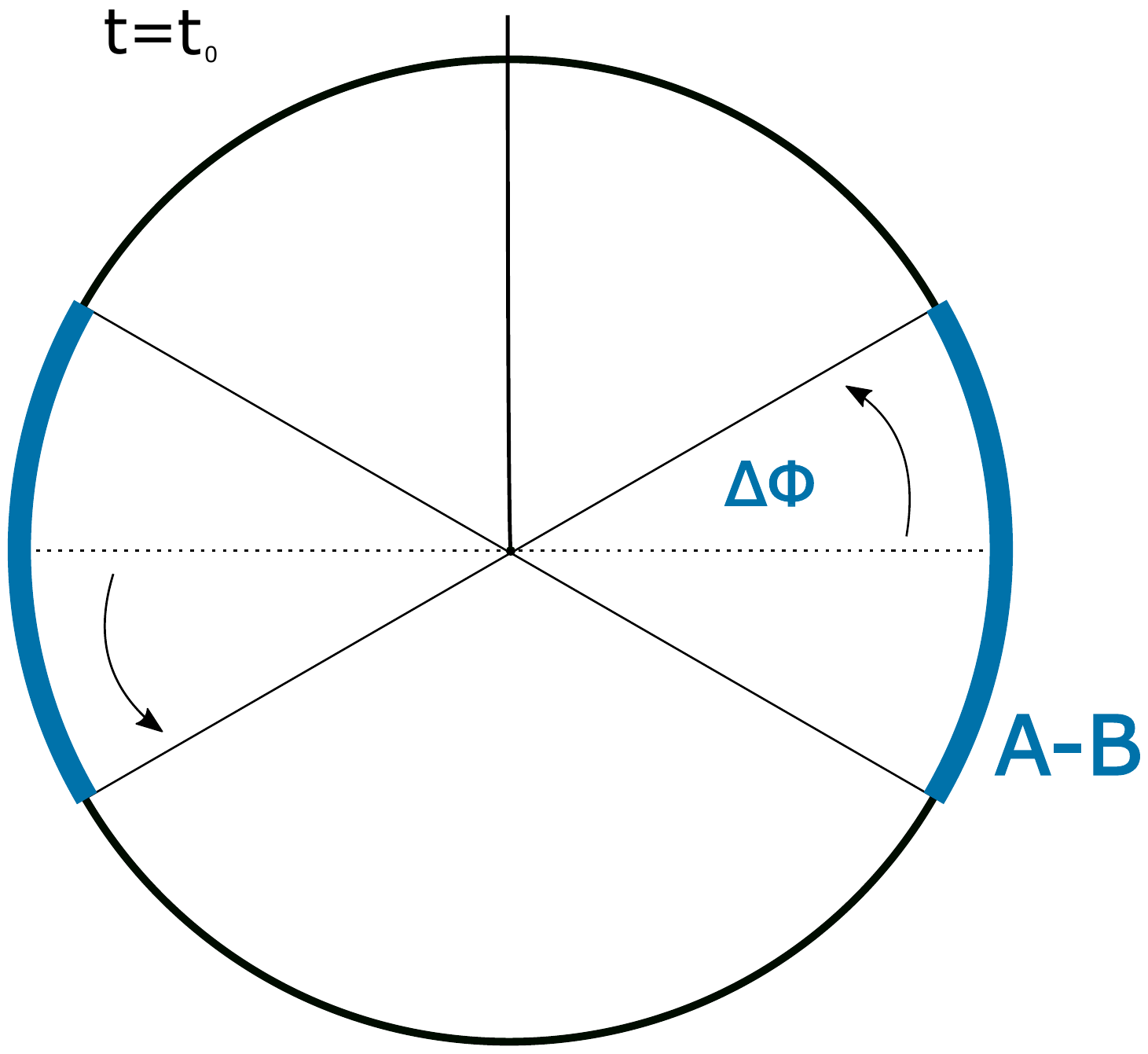}
\end{center}
\caption{\textbf{Diagram of the spatial region $A-B$ for a fixed time in one spatial dimension.}} \label{regions}
\end{figure}

We can use the liberty to choose the parameters $\phi _A$ and $\phi _B$ to simplify the form of the inequality. For instance, we can take $\phi _A$ and $\phi _B$ so that the second term in the right hand side of (\ref{eq:4.}) vanishes. This is achieved by considering $\phi _A=\pi /2 +  \Delta \phi$ and $\phi _B =\pi /2 -  \Delta \phi$ with $\Delta \phi \in (0,\pi /2)$. With this choice, we have
\begin{equation}\label{eq:5.}
 S_F^1(A,B) \le 2\pi R E \tan \left(\Delta \phi \right),
\end{equation}
where $E$ is the total energy of the state $\rho_1$, the integral over the whole space of the energy density. In figure \ref{regions} we sketch the region $A-B$ for a certain choice of $\Delta \phi$ in the case of one spatial dimension.

It is convenient to express the parameter $\Delta \phi$ in terms of the volume of the region $A-B$ and the total volume of the sphere $V_d$. This can easily be done by considering that
\begin{equation}\label{eq:8.}
V_{d}^{A-B}=
  \left(\frac{4 \Delta \phi}{2 \pi}\right)\frac{2\pi ^{\frac{d+1}{2}}}{\Gamma \left(\frac{d+1}{2}\right)}R^d=
  \left(\frac{4 \Delta \phi}{2 \pi}\right)V_d,
\end{equation}

With this in mind, inequality (\ref{eq:5.}) transforms into
\begin{equation}\label{eq:6.}
 S_F^1(A,B) \le 2\pi R  E \tan \left[\frac{\pi}{2}\left(\frac{V_d^{A-B}}{V_d}\right)\right].
\end{equation}

This last inequality sets an upper bound to the free entropy of the regions $A$ and $B$ for any state $\rho_1$. This bound depends on the total energy of the state, the region $A-B$ and the radius of the $d$ sphere $R$, which is just a parameter of the CFT.

In the limiting case in which the region $A-B$ is almost the total space, the bound does not say much, since in that case we would have $V^{A-B}_d/V_d\sim 1$ and therefore $S^1_f(A,B) \lesssim +\infty$. On the other hand, when the region $A-B$ is small compared with the size of the spatial slice of the cylinder, we would have $V_d^{A-B}/V_d \ll 1$ and therefore the free entropy of the state would go to zero. This is reasonable since the free entropy is monotonic with the size of the region $A-B$. Expanding the right hand side of (\ref{eq:6.}) for $V_d^{A-B}/V_d \ll 1$ and using (\ref{eq:8.}), we get
\begin{equation}
S_F^1(A,B) \le  E  V_d^{A-B} \left(\frac{\Gamma \left(\frac{d+1}{2}\right)}{2\pi^{\frac{d-3}{2}}R^{d-1}}\right)
 \left[1+
 \mathcal{O}\left(\frac{V_d^{A-B}}{V_d}\right)^2 \right]\,.
\end{equation}
Therefore, the first order term is linear in the energy of the state and the volume of the region $A-B$. The case of $d=1$ is particularly interesting, since the result is independent of the radius $R$
\begin{equation}\label{eq:9.}
 S_F^1(A,B) \le \frac{\pi}{2} E  L^{A-B}
 \left[1+
 \mathcal{O}\left(\frac{L^{A-B}}{L}\right)^2 \right]\,.
\end{equation}
In this case, the region $A-B$ corresponds to two infinitely separated segments in flat space.


\section{Discussion}\label{ch:discussion}

In this paper, we have used monotonicity of relative entropy to derive interesting relations between energy and entropy. For various CFT we have obtained interesting new quantum energy inequalities and energy-entropy relations that are analogous to the bound of Bekenstein.

The quantum energy inequalities we found for a two-dimensional CFT (\ref{eq:8}) and (\ref{cotita}) are generalisations of previous results. Interestingly, we have been able to re-derive (and improve for mixed states) the inequality by Fewster and Hollands through a different procedure, showing that there is a free entropy term which further restricts the possible amount of negative energy. We can conclude that as the entropy of a state increases, the possible amount of negative energy decreases, giving a clear relationship between two apparently disconnected features.

The fact that a purely local feature, such as the negative energy density, is constrained by a global quantity (the free entropy associated to some regions) is quite intriguing, though it may seem natural in view of the original motivation for negative energy constraints that assure the validity of the second law of thermodynamics \cite{ford}. An analogous situation ocurrs for the QNEC.

We have also found an interesting condition regarding the saturation of the property of monotonicity of relative entropy, finding a connection between this problem and the saturation of Fewster and Hollands inequality. This might be useful in order to further understand the physical significance of the saturation of the monotonicity.

Finally, the energy-entropy inequalities (\ref{bekelike}), (\ref{eq:17}) and (\ref{eq:4.}) constitute a new set of relations in which (a type of) entropy is bounded by some energy. This relations are well defined forms of Bekenstein-like bounds.

\section*{Acknowledgements}

The authors thank Guillem P\'erez-Nadal, Alan Garbarz and Gast\'on Giribet for discussions. This work was partially supported by CONICET, CNEA, Universidad Nacional de Buenos Aires, Instituto de Astronom\'ia y F\'isica del Espacio and Universidad Nacional de Cuyo, Argentina. H.C. acknowledges support from an It From Qubit grant of the Simons Foundation.


\begin{thebibliography}{99}
\bibitem{vedral}
  V.~Vedral,
  ``The role of relative entropy in quantum information theory'',
  Rev.\ Mod.\ Phys.\ {\bf 74} 197 (2002)
  [arXiv:quant-ph/0102094].

\bibitem{beke}
  H.~Casini,
  ``Relative entropy and the Bekenstein bound'',
  Class.\ Quant.\ Grav.\  {\bf 25}, 205021 (2008)
  [arXiv:0804.2182 [hep-th]].

\bibitem{bousso}
  R.~Bousso, H.~Casini, Z.~Fisher and J.~Maldacena,
  ``Entropy on a null surface for interacting quantum field theories and the Bousso bound'',
  Phys.\ Rev.\ D {\bf 91}, no. 8, 084030 (2015)
  doi:10.1103/PhysRevD.91.084030
  [arXiv:1406.4545 [hep-th]];
  R.~Bousso, H.~Casini, Z.~Fisher and J.~Maldacena,
  ``Proof of a Quantum Bousso Bound'',
  Phys.\ Rev.\ D {\bf 90}, no. 4, 044002 (2014)
  doi:10.1103/PhysRevD.90.044002
  [arXiv:1404.5635 [hep-th]].

\bibitem{first-law}
  D.~D.~Blanco, H.~Casini, L.~Y.~Hung and R.~C.~Myers,
  ``Relative Entropy and Holography'',
  JHEP {\bf 1308}, 060 (2013)
  doi:10.1007/JHEP08(2013)060
  [arXiv:1305.3182 [hep-th]].

\bibitem{negative}
  D.~D.~Blanco and H.~Casini,
  ``Localization of negative energy and the Bekenstein bound'',
  Phys. Rev. Lett. {\bf 111}, 221601 (2013)
  doi:10.1103/PhysRevLett.111.221601
  [arXiv:1309.1121 [hep-th]].

\bibitem{nega-necesaria}
  H.~Epstein, V.~Glaser and A.~Jaffe,
  ``Nonpositivity of energy density in Quantized field theories'',
  Nuovo Cim.\  {\bf 36}, 1016 (1965).

\bibitem{QEI's}
  C.~J.~Fewster,
  ``Lectures on quantum energy inequalities'',
  [arXiv:1208.5399v1 [gr-qc]].
  
\bibitem{leigh}
T.~Faulkner, R.~G.~Leigh, O.~Parrikar and H.~Wang, ``Modular Hamiltonians for
Deformed Half-Spaces and the Averaged Null Energy Condition'',
JHEP {\bf 1609}, 038 (2016)
doi:10.1007/JHEP09(2016)038
[arXiv:1605.08072 [hep-th]].\\
See also: T.~Hartman, S.~Kundu and A.~Tajdini, ``Averaged Null Energy Condition from Causality'', JHEP {\bf 1707} (2017) 066, [arXiv:1610.05308 [hep-th]].
 
\bibitem{wang}
S.~Balakrishnan, T.~Faulkner, Z.~U.~Khandker and H.~Wang, ``A General Proof of the Quantum Null Energy Condition'',
[arXiv:1706.09432 [hep-th]].

\bibitem{bekenstein}
  J.~D.~Bekenstein,
  ``A Universal Upper Bound on the Entropy to Energy Ratio for Bounded Systems'',
  Phys. Rev. D {\bf 23}, 287 (1981).

\bibitem{fewster}
  C.~Fewster and S.~Hollands,
  ``Quantum energy inequalities in two-dimensional conformal field theory'',
  Rev.\ Math.\ Phys.\ {\bf 17} (2005) 577-612.
  [arXiv:math-ph/0412028].

\bibitem{weak-mono}
  N.~Pippenger,
  IEEE Trans. Inf. Theory {\bf 49}, 773 (2003).

\bibitem{bisognano}
  J.~J.~Bisognano and E.~H.~Wichmann,
  ``On the Duality Condition for Quantum Fields'',
  J.\ Math.\ Phys.\  {\bf 17}, 303 (1976).
  doi:10.1063/1.522898.

\bibitem{teste}
  H.~Casini, E.~Teste and G.~Torroba,
  ``Modular Hamiltonians on the null plane and the Markov property of the vacuum state'',
  J.\ Phys.\ A {\bf 50}, no. 36, 364001 (2017)
  doi:10.1088/1751-8121/aa7eaa
  [arXiv:1703.10656 [hep-th]].

\bibitem{sphere}
  H.~Casini, M.~Huerta and Robert ~C.~Myers,
  ``Towards a derivation of holographic entanglement entropy'',
  JHEP {\bf 1105}, 036 (2011)
  doi:10.1007/JHEP05(2011)036
  [arXiv:1102.0440v2 [hep-th]].

\bibitem{cylinder}
  M.~Van~Raamsdonk,
  ``Lectures on Gravity and Entanglement'',
  {\it New Frontiers in Fields and Strings}. January 2017, 297-351.
  [arXiv:1609.00026v1 [hep-th]].
  
\bibitem{tonni}
  J.~Cardy and E.~Tonni,
  ``Entanglement hamiltonians in two-dimensional conformal field theory''
  [arXiv:1608.01283 [cond-mat.stat-mech]].
  
\bibitem{modular}
  R.~E.~Arias, D.~D.~Blanco, H.~Casini and M.~Huerta,
  ``Local temperatures and local terms in modular hamiltonians'',
  Phys. Rev. D {\bf 95}, 065005 (2017),
  doi:10.1103/PhysRevD.95.065005
  [arXiv:1611.08517 [hep-th]].
  
\bibitem{yngvason}
  H.~J.~Borchers and J.~Yngvason,
  ``Modular groups of quantum fields in thermal states'',
  J. Math. Phys. {\bf 40}, 601 (1999)
  [math-ph/9805013].
  
\bibitem{lashkari}
  N.~Lashkari, C.~Rabideau, P.~Sabella-Garnier and M.~Van~Raamsdonk,
  ``Inviolable energy conditions from entanglement inequalities'',
  JHEP {\bf 1506}, 067 (2015),
http://doi.org/10.1007/JHEP06(2015)067
[arXiv:1412.3514 [hep-th]].\\

%
  
\bibitem{monot1}
  D.~Sutter, M.~Tomamichel, A.~W.~Harrow,
  ``Strengthened Monotonicity of Relative Entropy
via Pinched Petz Recovery Map'',
  IEEE Transactions on Information Theory, vol. 62, no. 5, pages 2907-2913, 2016
  [arXiv:1507.00303v3].

\bibitem{difrancesco}
  P.~Di~Francesco, P.~Mathieu and D.~Senechal,
  ``Conformal field theory'',
  Springer, New York (1997).

\bibitem{cacardy}
  P.~Calabrese and J.~Cardy,
  ``Entanglement entropy and quantum field theory'',
  J.\ Stat.\ Mech.\ {\bf 0406}, P06002 (2004).
  [arXiv:hep-th/0405152].
  
\bibitem{ford} L. H. Ford, ``Quantum coherence effects and the second law of thermodynamics,'' Proc. Roy.
Soc. Lond. A 364 (1978). See also the discussion in \cite{negative}.









\end{thebibliography}
\end{document}